\documentclass{article}

\usepackage[intlimits]{amsmath}
\usepackage{amssymb}
\usepackage{graphicx}
\usepackage{url}
\usepackage[unicode]{hyperref}
\usepackage[all]{hypcap} 

\newcommand{\ph}{\varphi}
\newcommand{\rh}{\varrho}
\newcommand{\nn}{\nonumber\\}
\newcommand{\exv}[1]{\left\langle{#1}\right\rangle}
\newcommand{\exvs}[1]{\langle{#1}\rangle}
\renewcommand{\d}{\partial}

\title{Shear viscosity of the $\Phi^4$ theory from classical
  simulation}
\author{M. M. Homor and A. Jakovac}
\date{\today}

\begin{document}

\maketitle

\begin{abstract}
  Shear viscosity of the classical $\Phi^4$ theory is measured using
  classical microcanonical simulation. To calculate the Kubo formula,
  we measure the energy-momentum tensor correlation function, and
  apply the Green-Kubo relation. Being a classical theory, the results
  depend on the cutoff which should be chosen in the range of the
  temperature. Comparison with experimentally accessible systems is
  also performed.
\end{abstract}

\section{Introduction}

Transport coefficients, in particular shear viscosity, are very hardly
accessible quantities in perturbative quantum field theory
calculations. Transport is characteristic to systems where information
is spread by diffusion, the time evolution is $\sim\sqrt{Dt}$, the
diffusion constant being the corresponding transport coefficient. The
diffusion constant itself is proportional to the quasiparticle
lifetime $D\sim \tau$. This is infinite in a free gas, and inversely
proportional to some powers of the coupling constant for weak
couplings. Therefore the perturbative evaluation of the Kubo formula
\cite{Hosoya:1983id} requires resummation of an infinite set of
diagrams \cite{Jeon:1994if}. To circumvent this difficulty one can use
effective methods to calculate the transport coefficients. One of
these methods is the use of Boltzmann equations which is equivalent
with the resummation of the singular part of the full perturbation
series \cite{Jeon:1995zm,Jakovac:2001kj}. Boltzmann equation method is
used to obtain general results in gauge theories
\cite{Arnold:2000dr,Arnold:2003zc} proving that to the leading order
one has a shear viscosity $\eta\sim \frac1{g^4\ln g}$. Boltzmann
equation methods are used also in other models to compute shear
viscosity, like in meson models
\cite{Dobado:2003wr,Buballa:2012hb,Mitra:2013gya} or in full QCD
\cite{Xu:2004mz, Xu:2007aa}. Other perturbation theory motivated
methods to calculate the shear viscosity are 2PI resummation techniques
\cite{Aarts:2004sd} or the generalized quasiparticle approach
\cite{Peshier:2005pp}.

Apart from the technical difficulties, also the applicability of
perturbation theory makes these results less relevant for strongly
interacting QCD-like systems. Small value of the shear viscosity of
the QCD plasma, reported by analyses of experimental data
\cite{Adler:2003kt} suggests that the QCD matter is close to a perfect
liquid \cite{Liao:2009gb}. This implies that the interaction is rather
strong, the quasiparticle lifetime is very short, and so perturbation
theory is hardly applicable.

Where perturbation theory is not well applicable, one seeks
non-perturbative methods. Computer Monte Carlo (MC) simulation of QCD
was used to extract shear viscosity data roughly in agreement with
measurements \cite{Meyer:2007ic}. The temporal range of the Euclidean
formalism of the MC setup, however, makes the correlations less
sensitive to long range physics which are relevant for transport
\cite{Petreczky:2007js}. Other popular method is to use the dual
theory approach, based on AdS/CFT correspondence. Then weakly coupled
five dimensional gravity can be used to compute transport coefficients
in strongly coupled (conformal) field theories \cite{Kovtun:2003wp,
  Kovtun:2004de}. There are several model studies in this field which
calculate shear viscosity by this method.

Another nonperturbative method to approach the dynamics of quantum
field theory is the use of classical theories to study both
equilibrium
\cite{Aarts:1996qi,Nauta:1997yi,Buchmuller:1997yw,Buchmuller:1997nw,Aarts:2001yn,Holdom:2006tt}
and nonequilibrium phenomena
\cite{Borsanyi:2002tm,Borsanyi:2003ib,Boyanovsky:2003tc,Berges:2014yta}.
Here one applies classical equations of motion starting from some
initial conditions, and solve them by numerical methods on a finite
mesh. The system thermalizes\footnote{Note that we work with finite
  systems with finite energy density where thermalization is
  possible.}, which in a classical system means equipartition of the
energy. From the classical trajectories we can evaluate expectation
values of different observables as time averages.

From the point of view of perturbation theory classical and quantum
systems are similar \cite{Buchmuller:1997yw}. In particular one can
study expectation values of composite operators like
$\left\langle\Phi^2(x)\Phi^2(y)\right\rangle$. Comparing the classical
and quantum computations one finds that with an appropriate choice of
the cutoff of the classical theory $\Lambda_{cl}\sim T$ the quantum
results can be nicely reproduced \cite{Jakovac:1998jd}.

Encouraged by these results we tried to use classical simulations to
compute the shear viscosity in a simple bosonic classical system, the
$\Phi^4$ model. The Kubo formula for the shear viscosity
\cite{Hosoya:1983id} contains commutator of spatial components of the
energy-momentum tensor. We computed it with help of the Green-Kubo
relation which is the classical counterpart of the quantum
Kubo-Martin-Schwinger relation (fluctuation dissipation theorem)
\cite{LeBellac}. This system has the potential to show a phase
transition, similarly to the QCD case (although it is a second order
here, as opposed to the crossover nature in QCD). This makes possible
to study the $\eta/s$ ratio near the phase transition.

The paper is organized as follows. First we overview the details of the
discretization and classical simulation method for the $\Phi^4$
model in Section II. In Section III we discuss the thermalization
process and the measured characteristics of the thermal equilibrium,
in particular thermal mass. In Section IV we report on our results of
the energy-momentum tensor correlation functions and the classical
values of the shear viscosity. In Section V we apply our method to
quantum systems and present the $\eta/s$ ratio, also in comparison
with the experimentally measured values in different systems. The paper
is closed with Conclusions.

\section{The system, discretization and simulation algorithm}

The system we study is the quartic scalar model, which has the
Hamiltonian density
\begin{equation}
  \label{eq:Hamiltonian}
  \mathcal{H}_{\mathbf{x}} =  \frac12 \Pi^2(\mathbf{x}) +
  \frac12(\nabla \Phi(\mathbf{x}))^2 +
  \frac{m^2}2\Phi^2(\mathbf{x}) +\frac{\lambda}{24}\Phi^4(\mathbf{x}).
\end{equation}
Here $\Phi$ denotes the field, $\Pi$ its canonical conjugate. The
corresponding equations of motion (EoM) are
\begin{equation}
  \label{eq:EoM}
  \dot \Phi = \Pi,\qquad \dot\Pi= \triangle \Phi -m^2 \Phi
  -\frac{\lambda}6\Phi^3.
\end{equation}
We remark that by rescaling the fields $\Phi\to\Phi/\sqrt{\lambda}$
and $\Pi\to\Pi/\sqrt{\lambda}$, the equations of motion become
$\lambda$-independent. We could work therefore with $\lambda=1$, but
for better readability we keep the notation of $\lambda$.

We discretize the model on a symmetric finite spacelike mesh 
\[U=\{ {\bf x}= \sum_{i=1}^3 n_ia {\bf e}_i \,|\, n_i= 0\dots N-1\},
\]
where ${\bf e}_i$ are orthogonal unit vectors and $a$ is the lattice
spacing; we express all dimensional quantities in lattice units and so
we choose $a=1$. For the lattice size we have in our simulations
$N=36,\,40$ and $50$, and use periodic boundary conditions. The
discretized Laplacian is
\[
\triangle\Phi(\mathbf{x})=\sum_{i=1}^3\left[\Phi
  (\mathbf{x}+\mathbf{e}_i)-2\Phi(\mathbf{x})+\Phi(\mathbf{x}-\mathbf{e}_i)
  \right].
\]
The discretized Hamiltonian can be written as $H=\sum_{\mathbf{x}\in U}
\mathcal{H}_{\mathbf{x}}$, where the Hamiltonian density formally
equivalent to \eqref{eq:Hamiltonian}, with $(\nabla\Phi(\mathbf{x}))^2
= \sum_{i=1}^3 \left[ \Phi(\mathbf{x}+\mathbf{e}_i) -
  \Phi(\mathbf{x})\right]^2.$ This is, however, not a local expression
anymore, as it connects nearest neighbor field values.

For the evaluation of expectation values we also need Fourier
transformation. It is defined on the reciprocal lattice $\bar{U}$ with
the following definition (which corresponds to the \texttt{fftw++}
conventions \cite{fftw})
\begin{equation}
  f_{\textbf{k} \in \bar{U}}=\sum_{\textbf{x}\in U} \exp^{-2\pi i(\textbf{kx})/N} 
  f_{\textbf{x}}, \qquad
  f_{\textbf{x} \in {U}}=\frac1{N^3}\sum_{\textbf{k}\in \bar U} \exp^{2\pi i(\textbf{kx})/N}
  f_{\textbf{k}}.
\end{equation}
The reciprocal lattice is equivalent with the original lattice in case
of cubic lattices we used. The Fourier-transformed Hamiltonian reads
\begin{equation}
  H= \frac1{N^3}\sum_{\mathbf{k}\in\bar U} \left[\frac12 \vert
    \Pi_{\textbf{k}}\vert^2 + \frac12 \omega_k^2 \vert
    \Phi_{\textbf{k}}\vert ^2\right] + 
  \frac{\lambda}{24N^6}\!\!\!\sum_{\mathbf{k}_i\in\bar
    U}\!\!\!\Phi_{\textbf{k}_1} \Phi_{\textbf{k}_2}\! 
  \Phi_{\textbf{k}_3}\! \Phi_{\textbf{k}_4}.
  \label{eq:diszkrHam}
\end{equation}
where $\omega_k^2 = m^2+\sum_{i=1}^3 4\sin^2 \left(
  \frac{\pi\textbf{ke}_i}{N} \right)$, and $\sum_i\mathbf{k}_i=0$ in
the last term.

The time evolution in computer is realized using the leap-frog
algorithm. Here one chooses a time step $dt$, so at $n$th step one
arrives at time $t=n\,dt$. In the time step from $n-1$ to $n$, the two
equations of \eqref{eq:EoM} are treated subsequently: first one
evolves the field configuration
\begin{equation}
\Phi^{(n)}(\mathbf{x})=\Phi^{(n-1)} + dt\,\Pi^{(n-1)}(\mathbf{x}),
\end{equation}
then the canonically conjugated field configuration, using the new
values of the field:
\begin{equation}
\Pi^{(n)}(\mathbf{x})=\Pi^{(n-1)}(\mathbf{x})+dt\, \left(
  \triangle\Phi^{(n)}(\mathbf{x})-m^2 \Phi^{(n)}(\mathbf{x}) -
  \frac{\lambda}{6}\left(\Phi^{(n)}(\mathbf{x})\right)^3\right),
\end{equation}
with the discretized Laplacian. 

We can use the notion of the energy in the discretized model, too, as
$E=H=\sum_{\mathbf{x}\in U} \mathcal{H}$. This quantity is conserved
only for continuous time evolution; since we evolve the time in
discrete steps, the total energy is not necessarily conserved. An
important consistency check for the reliability of the algorithm is
that in a long run the energy remains conserved. The leap-frog
algorithm satisfies this requirement.

The classical ground state of the system, ie. the minimum of the
energy is at spatially homogeneous field. If $m^2$ and $\lambda$ is
positive, the minimum is reached at $\Phi=\Pi=0$. The coupling
$\lambda$ must be positive otherwise the Hamiltonian of the system is
not bounded from below. If $m^2<0$, the minimal energy is reached at a
finite $|\Phi|=\Phi_0$ value: this is the spontaneous symmetry broken
(SSB) phase. The minimum condition yields $\Phi_0^2=
\frac{-6m^2}\lambda$.

\section{Description of the thermal equilibrium}

Since we are primarily interested in the equilibrium properties of the
system, we may start from an arbitrary initial condition. Practically
we started with $\Phi(\mathbf{x}) \equiv 0$, and from random values of
$\Pi(\mathbf{x})$. After a certain time evolution (practically around
$t/a\sim 10000$) we arrive at a steady equilibrium state. While the
complete system forms a microcanonical ensemble, for local observables
we can use a canonical ensemble. To determine its properties, we have
taken the histogram of $\Pi_{\mathbf{x}}$, and found that it can be
described by a Gaussian. This corresponds to the Boltzmann
distribution $\sim e^{-\beta\Pi_x^2}$ with $\beta$ as a parameter
interpreted as the ``inverse temperature''. Therefore one can compute
the expectation value of a local operator $A(\Phi,\Pi)$ which depends
on the fields as
\begin{equation}
\left\langle A(\Phi,\Pi) \right\rangle =
\frac1Z \int \prod_{x\in U} d\Phi_xd\Pi_x A(\Phi,\Pi) e^{-\beta H(\Phi,\Pi)},
\end{equation}
where $Z=\int \prod_{x\in U} d\Phi_xd\Pi_x e^{-\beta H(\Phi,\Pi)}$.
In Fourier space we should handle the problem that there is a relation
between the integration variables $\Phi_{\mathbf{k}} =
\Phi_{-\mathbf{k}}^*$, and similarly for $\Pi_{\mathbf{k}}$, because
of $\Phi$ and $\Pi$ in coordinate space are real. To overcome this
problem we introduce a purely real field
\begin{equation}
\tilde{\Phi}_{\textbf{k}}=\left\lbrace 
\begin{array}{l l}
\mathrm{Re}\Phi_{\textbf{k}} & \textrm{if } k_3 \geq 0 \\
\mathrm{Im}\Phi_{\textbf{k}} & \textrm{if } k_3 < 0. \\
\end{array}\right. 
\end{equation}
This allows to write
\begin{equation}
  \langle A(\Phi,\Pi)\rangle =
  \frac1Z\int \prod_{\textbf{k} \in \bar{U}} d\tilde\Pi_{\textbf{k}}
  d\tilde \Phi_{\textbf{k}}\, A(\Phi,\Pi) e^{-\beta H(\Phi,\Pi)}. 
\end{equation}

\begin{figure}[htbp]
  \centering
  \includegraphics[scale=0.7]{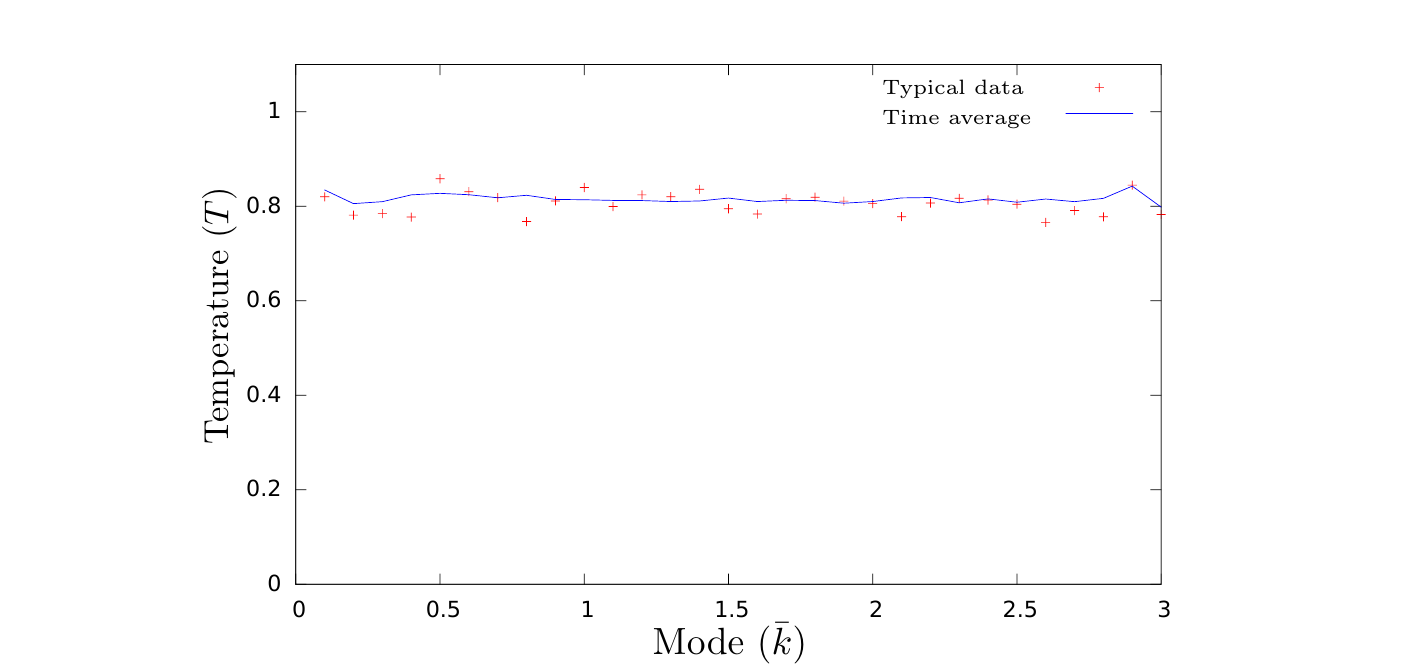}
   \caption{Mode dependence of temperature ($\bar{k}^2=\sum_{i=1}^3 \sin ^2 \left( 2 \pi \frac{\textbf{ke}_i}{N} \right) $)}
   \label{fig:ekvipart1}
\end{figure}

To measure the temperature we use the relation
\begin{equation}
  \langle\vert \Pi_{\textbf{k}}\vert^2\rangle =
  \frac{1}{\tilde{Z}_{\textbf{k}}} \int_{-\infty}^{\infty} d\tilde{\Pi}_{\textbf{k}}
  \left( \tilde{\Pi}_{\textbf{k}}^2 + \tilde{\Pi}_{-\textbf{k}}^2\right) 
  e^{-\beta \frac{1}{2N^3}\left(\tilde{\Pi}_{\textbf{k}}^2 +
      \tilde{\Pi}_{-\textbf{k}}^2\right)}= 2N^3 T
\end{equation}
with $T=1/\beta$. Using this formula we can check that the system
arrived at equilibrium, by verifying that $\langle\vert
\Pi_{\textbf{k}}\vert^2\rangle$ is independent of $\mathbf{k}$
(equipartition). An example for this distribution in a completely
thermalized state is shown in Fig.~\ref{fig:ekvipart1}.

We remark that the above canonical equilibrium description is in fact
a 3D field theory of the initial conditions. As compared to the
original action which is four dimensional, we have a dimensionally
reduced theory. As a consequence the mass (energy) dimension of the
field is $[\Phi]=1/2$. This fact will be used later when we apply
dimensional analysis.

In the thermal equilibrium we can perform perturbation
theory. Although at large coupling (where we actually performed our
simulations) results of perturbation theory are not necessarily
perfect, but in several aspects these may ``guide the eye'' to
understand some robust features of the results. The details of
perturbation theory in the classical theory can be found in
\cite{Buchmuller:1997yw}. One uses here two types of propagators
\begin{equation}
  G_{ret}(k) = \frac1{(k_0+i\varepsilon)^2 -\omega_k^2},\qquad
  iG_{3D}(k) = \frac T{k_0} \,\rh(k)
\end{equation}
where the free spectral function is
\begin{equation}
  \label{eq:freespect}
  \rh(k) = 2\pi \mathrm{sgn}(k_0)\delta(k_0^2-\omega_k^2),
\end{equation}
and
\begin{equation}
  \omega_k^2 = \sum_{i=1}^3 4\sin ^2 \left(\frac{\pi
        k_i}{N}\right) + m^2,\qquad k_i\in 0\dots N-1.
\end{equation}

As a final, technical issue, we remark that solving the field equation
corresponds to the pure microcanonical, energy conserving approach to
the thermodynamics. However, knowing that the system reaches
equilibrium with Boltzmann distribution, we can also use a canonical
approach with a heat bath: in the language of the equations of motion
it can be realized as a Langevin equation. There we introduce a
$\gamma$ damping parameter and a noise represented by
$\xi(\mathbf{x})$ independent stochastic variables with uniform
distribution at each time step. We then change the update of $\Pi$ to
\begin{equation}
\Pi^{(n)}(\mathbf{x})=(1-\gamma dt) \Pi^{(n-1)}(\mathbf{x})+dt \left( \triangle\Phi^{(n)}(\mathbf{x})-m^2 \Phi^{(n)}(\mathbf{x})-\frac{\lambda}{6}\left(\Phi^{(n)}(\mathbf{x})\right)^3+\xi(\mathbf{x})\right).
\end{equation}
This stochastic process drives the system towards an equilibrium
distribution with ${\cal P}(E) \sim e^{-\beta E}$ distribution
function. Because of the Einstein relation $2\gamma T = \exvs{\xi\xi}$
we can control the temperature of the thermal distribution. This
algorithm can largely speed up the thermalization.  After the system
arrived at the equilibrium, we switched off the noise and damping terms
in order that it does not influence the measurements.

\section{Equilibrium observables}

After we reached the equilibrium state, we can measure expectation
values using time average
\begin{equation}
  \left\langle A(\Phi,\Pi)\right\rangle =
  \frac1t\int\limits_{t_0}^{t_0+t}\! dt' A(\Phi(t),\Pi(t)).
\end{equation}
The equilibrium system can be characterized by a single value, for
example the temperature.

\subsection{Energy}

\begin{figure}[htbp]
  \centering
  \includegraphics[scale=0.7]{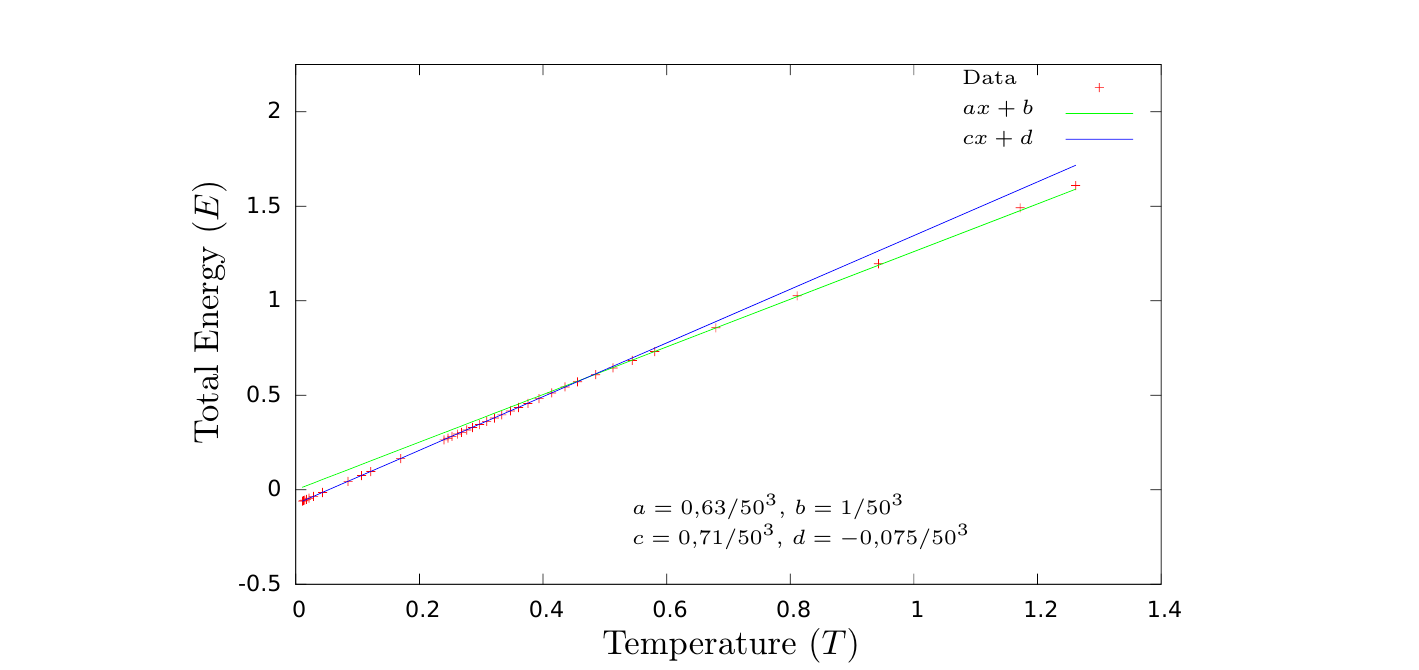}
  \caption{Temperature dependence of total energy ($N^3=V=50^3$), $\lambda=5$, $m^2=-0.5$}
  \label{fig:ETFitT50L5m05}
\end{figure}

We measured the relation between the temperature and the energy
density, the results are shown in Fig.~\ref{fig:ETFitT50L5m05}.. We
found that the relation is linear, with slightly different slope in
the symmetric and SSB regimes.
One can clearly identify the phase transition region. Since our goal
was not to study the phase transition point very accurately, we did not
try to focus on this regime close enough to be able to tell details
about it.

This figure tells us that the heat capacity is proportional to the
number of modes, as it is expected from a classical theory. Rewriting
the lattice spacing $a$, this also means that the specific heat is
proportional to $a^3$. This is the well known Rayleigh instability
(ultraviolet catastrophe) of the classical plasma. To have physically
meaningful result, the lattice spacing must have a finite value.

\subsection{Mass and symmetry breaking}

It is important to notice, that the mass parameter of the Lagrangian
(the bare mass) is not the same as the mass appearing in the observables
(the effective mass). Physically it happens because of the nontrivial
effect of the fluctuations. 

To estimate this effect (cf. Refs.\@ \cite{Borsanyi:2002tm,Borsanyi:2003ib})
we used background field method. We shifted the classical field with
its expectation value: $\Phi\to\Phi_0+\ph$, where
$\left\langle\ph\right\rangle=0$. The shifted Lagrangian reads
\begin{eqnarray}
  {\cal L}=&&-\frac{m^2}2\bar\Phi^2 -\frac\lambda{24}\bar\Phi^4 
  -\ph\left(m^2\bar\Phi +\frac\lambda6\bar\Phi^3\right) +\frac12
  \ph(-d^2-m^2)\ph - \frac\lambda4 \bar\Phi^2\ph^2 -\nn&&- \frac\lambda6
  \bar\Phi\ph^3 - \frac\lambda{24} \ph^4. 
\end{eqnarray}
To lowest order (Hartree approximation) we substitute the fluctuations
by their expectation values. Using the fact that $\exv{\ph}=0$ we find
up to a constant
\begin{equation}
  {\cal L}=-\frac12\left(m^2 + \frac\lambda2\exv{\ph^2}\right)
  \bar\Phi^2 -\frac\lambda{24}\bar\Phi^4.
\end{equation}
This means that the effective mass is modified by the effect of the
fluctuations. Since the mass dimension of the field is $[\ph]= {1/2}$,
by dimensional reasons $\exv{\ph^2}\sim T$. On the other hand this is
a correction to the mass squared, and so the coefficient is also
dimensionful, with finite lattice spacing it is proportional to
$a^{-1}$. The coefficient in leading order in perturbation theory
reads
\begin{equation}
  \exv{\ph^2} = \frac T{N^3} \sum\limits_{k\in\bar U}
  \frac1{ \sum_{i=1}^3 4\sin ^2 \left(\frac{\pi
        k_i}{N}\right) + m^2} \stackrel{N\to\infty}{\longrightarrow}
  0.2527 T  - \frac{mT}{4\pi} + {\cal O}(m^2),
\end{equation}
but this number is unreliable for large couplings.

\begin{figure}[htbp]
  \centering
  \includegraphics[scale=0.7]{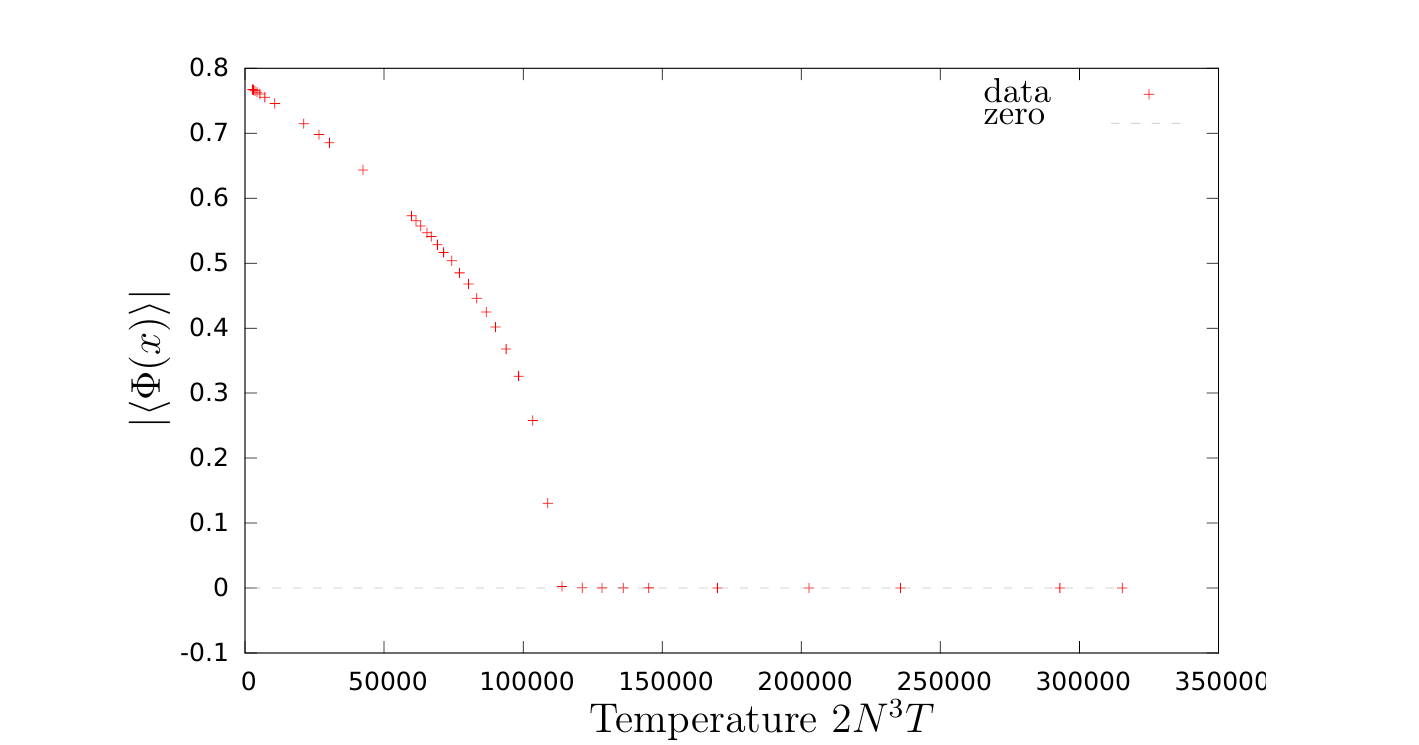}
  \caption{Temperature dependence of the expectation value of the field.}
  \label{fig:ssb}
\end{figure}

One consequence of this formula is that at fixed negative bare mass
the effective mass term will be positive at high enough
temperature. This means that the minimum of the effective action for
the constant field (the constrained free energy) will become zero at
this temperature: the symmetry is restored. To see it we measured the
expectation value of the field for various tree level masses at
different temperatures. The result can be seen on Fig.~\ref{fig:ssb}.

A more delicate question is that in the 3D classical field theory,
unlike in the 4-dimensional theory, the temperature influences the
renormalization. This means that the meaning of mass and temperature
cannot be separated from each other in that clear way as it can be
done in the 4D case. We therefore also performed simulations with
fixed effective mass at different temperatures: this requires to tune
the bare mass parameter. In practice we fix the desired effective mass
and the bare mass, and tune the temperature accordingly with the
application of Langevin equations described earlier.

\begin{figure}[htbp]
  \centering
  \hspace*{-1cm}\includegraphics[scale=1.0]{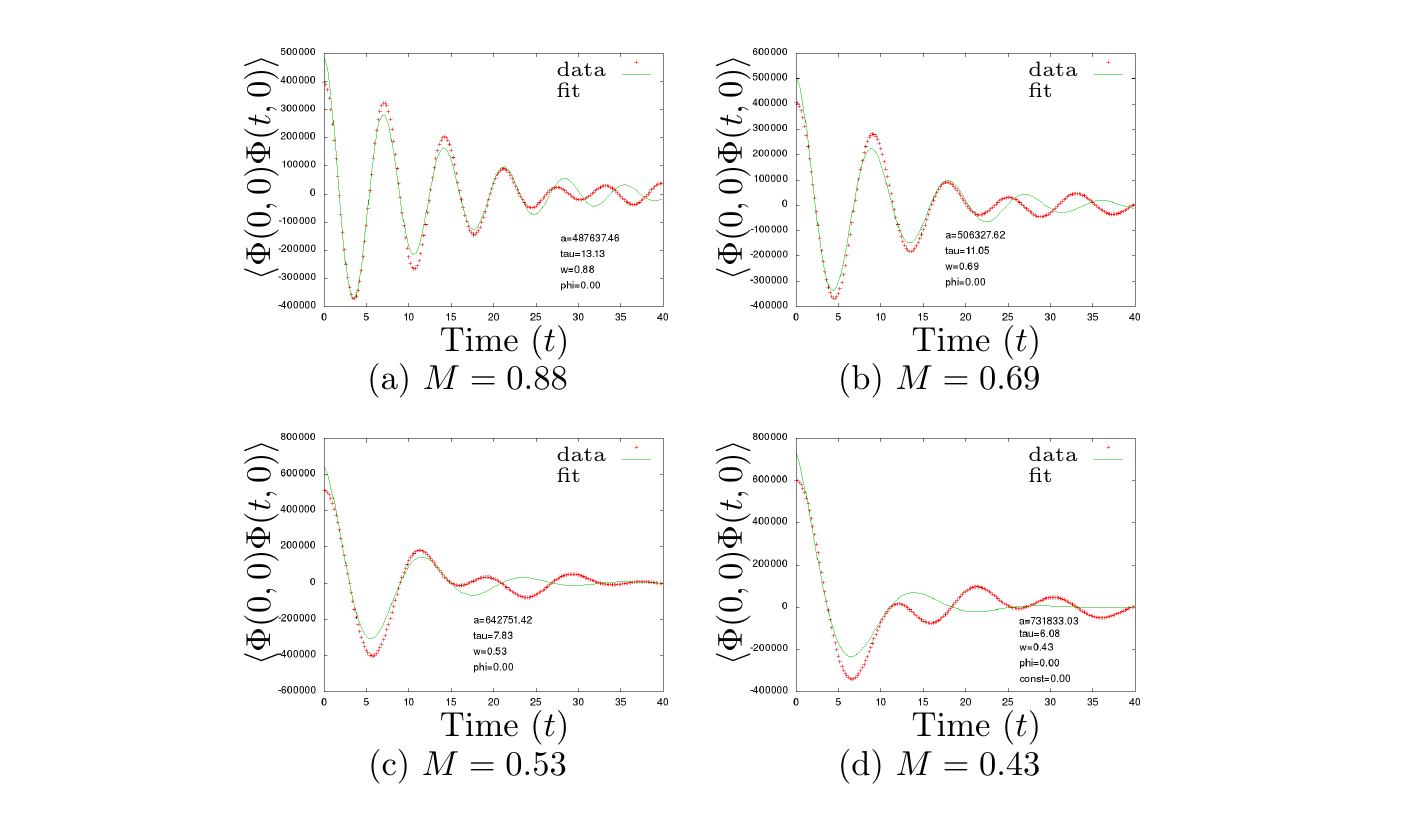}
  \caption{The real time behavior of the field correlation function
    at different bare masses, $\lambda = 5$.}
  \label{fig:mGTK}
\end{figure}

For the definition of the mass we measured the correlation function
\begin{equation}
G(t,\textbf{x}):=\langle\Phi(t,\textbf{x})\Phi(0)\rangle
\end{equation}
In leading order of perturbation theory we expect that this correlator
is the free one, where we should also take into account the mass
modification:
\begin{equation}
G(t,\textbf{k})=\left\langle |\Phi_k|^2\right\rangle \cos \omega_k t,
\end{equation}
where $\omega_k^2=\textbf{k}^2+m^2+\frac\lambda2\exv{\ph^2}$. If one
goes beyond the first order of perturbation theory, then one obtains
self-energy corrections, and the pure harmonic behavior of the
correlator will be spoiled. If a one-particle particle mass shell is
dominant, then we can speak about quasiparticle excitations. In that
case in real time evolution one can observe a damped oscillation
\begin{equation}
  G(t,\textbf{k}) \sim \exp(-t/\tau_k)\cos(\omega_kt).
\end{equation}
However, the closer we are to the phase transition point, the worse
behavior could be observed for the static $\Phi$ field, as it is
demonstrated in Fig.~\ref{fig:mGTK}.
We can see that far from the phase transition point, where the
effective mass is large, the quasiparticle assumption is valid. With
decreasing effective mass the fit works worse and worse. In this case
the definition of the notion ``mass'' is not unique anymore. For a
more sophisticated description we should use the complete spectral
function; but we just need a characterization of the mass and
temperature. For that purpose we use the best quasiparticle fit to the
real time data for the zero mode. This is some mean value of the
spectral peak, in the vicinity of the phase transition point it
remains finite, as opposed to the inverse spatial correlation length.

\begin{figure}[htbp]
  \centering
  \includegraphics[scale=0.6]{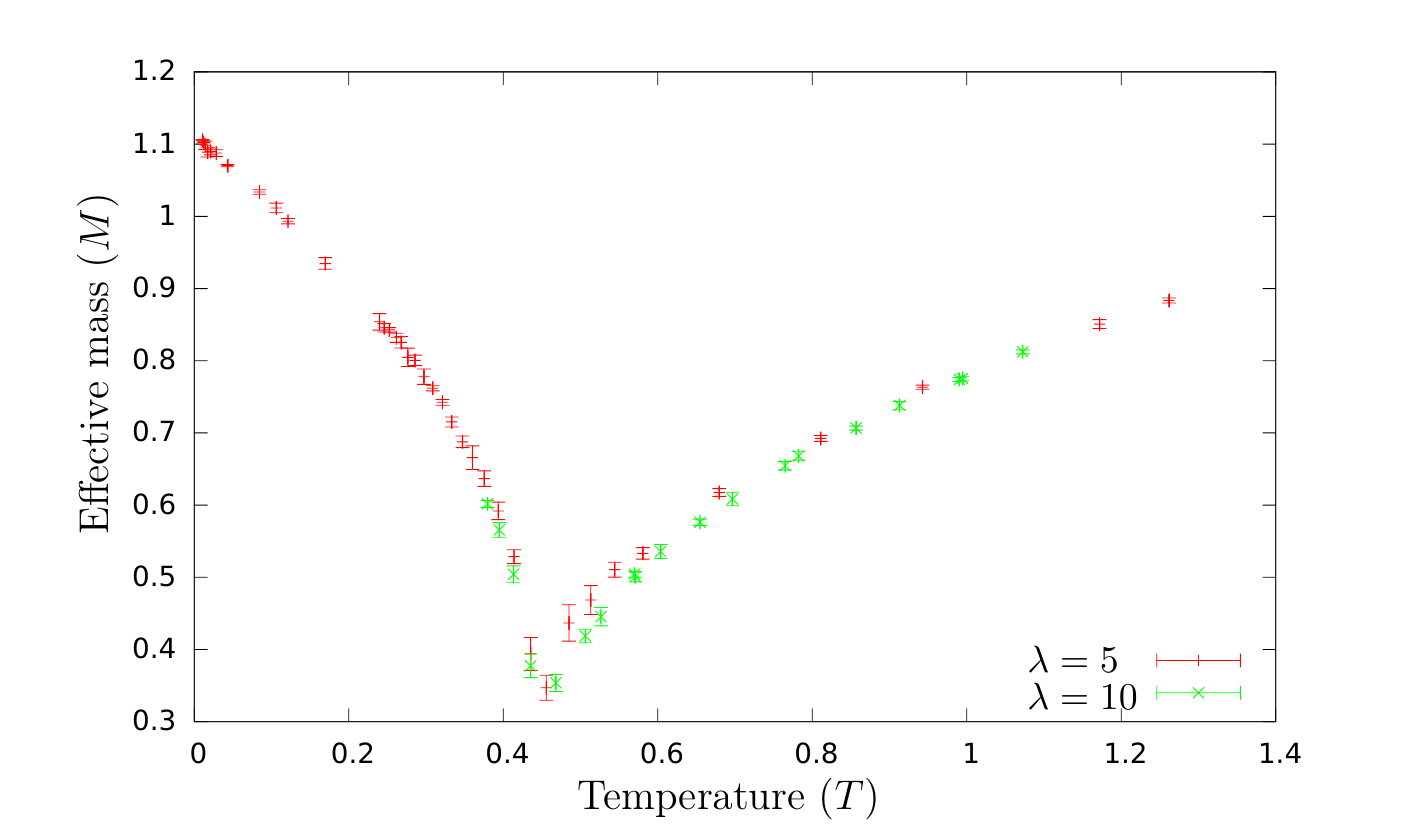}
  \caption{Temperature dependence of the effective mass parameter for
    various interaction strength ($\lambda$) where $m^2=-0.5$. It has
    a minimum at the phase transition point, but with our definition
    the minimum is not at zero.}
  \label{fig:MTdiffL}
\end{figure}

\begin{figure}[htbp]
  \centering
  \hspace*{-1cm}\includegraphics[scale=1.0]{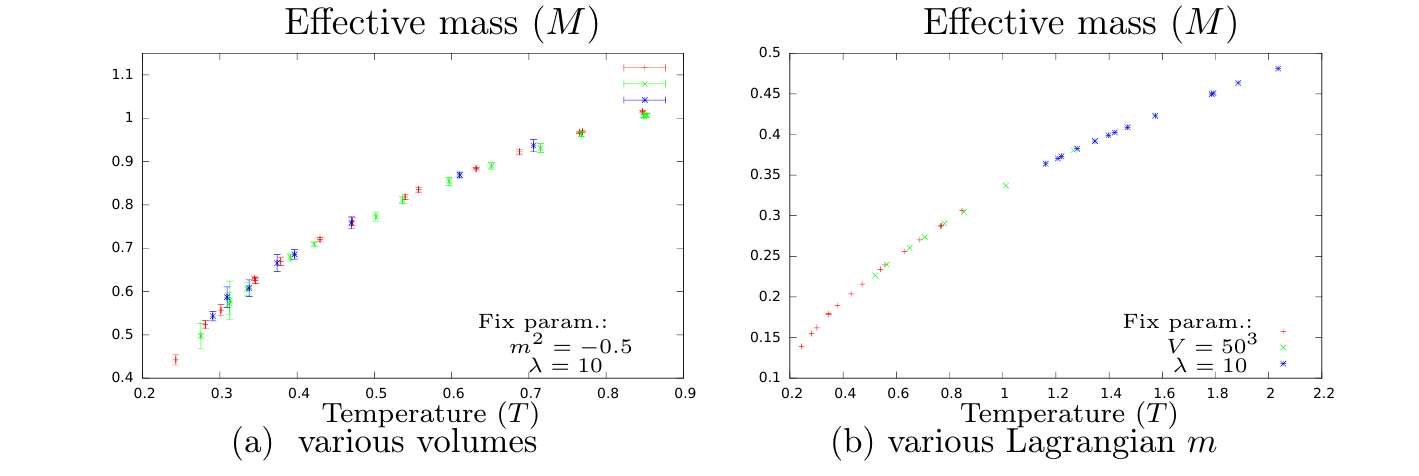}
  \caption{Left panel: effective mass at different volumes. Right panel: effective mass in theories with different bare mass.}
  \label{fig:tomeg}
\end{figure}

The temperature dependence of the so-defined mass is shown in
Fig.~\ref{fig:MTdiffL} with fixed bare mass $m^2=-0.5$. We can clearly
see the position of the phase transition point which sits at the
minimum of this curve.
One may also check, whether we reached the infinite volume
(thermodynamical) limit. For that we determined the temperature
dependence of the mass at various volumes, see left panel of
Fig.~\ref{fig:tomeg}. This plot suggests that we reached already the
thermodynamical limit. We can also check the temperature dependence of
the effective mass, this is shown in the right panel of
Fig.~\ref{fig:tomeg}. We see that for different bare masses the
effective mass values sit on a unique curve.

\section{Viscosity}

The central topic of this paper is the determination of the shear
viscosity. The Kubo-formula for momentum transport
\cite{Hosoya:1983id} requires to compute
\begin{equation}
  \eta=\lim_{\omega \rightarrow 0}\frac{\rh_{T_{12}T_{21}}
    (\omega,\textbf{k}=0)}{\omega},
\end{equation}
where
\begin{equation}
  \rh_{AB}(x) = \langle \left[ A(x),B(0)\right] \rangle,
\end{equation}
$\eta$ is the shear-viscosity and $T_{12}$ is the $12$ component of
the energy momentum tensor; in case of the scalar field theory it is
$T_{12}=\d_1 \Phi \d_2 \Phi$. In classical theory we cannot measure
the commutator of two operators, but we can measure the correlation
function instead. For $A(x)$ and $B(0)$ operators it is defined as
\begin{equation}
  S_{AB}(x) = \langle A(x) B(0)\rangle_{cl},
\end{equation}
where the ``cl'' subscript refers to the classical correlation
function. To connect this quantity with the viscosity we use the
Green-Kubo formula, which claims that
\begin{equation}
  \rh_{AB,cl}(\omega,\mathbf{k}) = \beta\omega S_{AB,cl}(\omega,\mathbf{k}),
\end{equation}
which is a direct consequence of the quantum Kubo-Martin-Schwinger
relation
\begin{equation}
  \rh_{AB}(\omega,\mathbf{k}) = (1-e^{-\beta\omega})S_{AB,cl}(\omega,\mathbf{k}),
\end{equation}
in the $\beta\omega\to0$ limit. Using this relation the viscosity is
\begin{equation}
  \label{eq:GreenKubo}
  \eta_{cl} = \beta S_{T_{12}T_{12}}(k=0).
\end{equation}

\begin{figure}[htbp]
  \centering
  \includegraphics[scale=0.75]{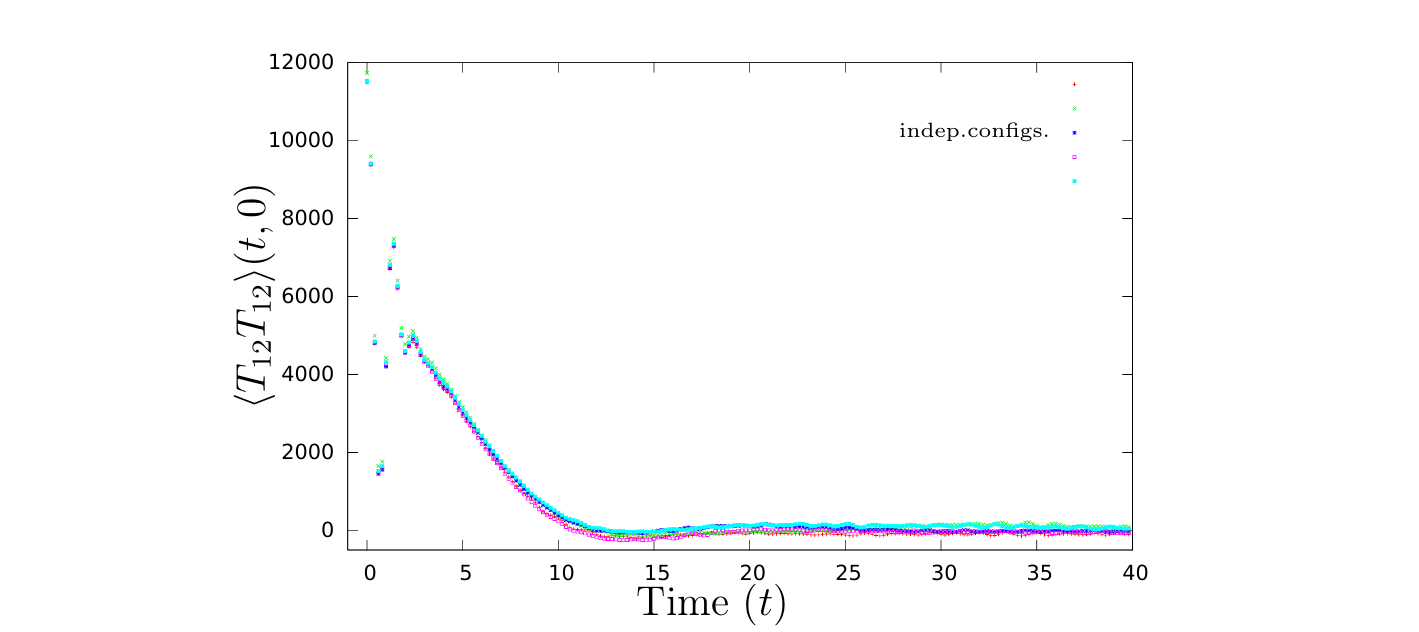}
  \caption{$T_{12}$ autocorrelation function in real time.}
  \label{fig:viszkot}
\end{figure}

The direct result of our simulations in real time can be seen in
Fig.~\ref{fig:viszkot}. We repeated the measurements of
$S_{T_{12}T_{12}}$ for five different configurations, meaning that
after each measurement we allowed the system evolve in time until we
reached an independent configuration. This makes it possible to
estimate the statistical error of the simulation.

\begin{figure}[htbp]
  \centering
  \includegraphics[scale=0.75]{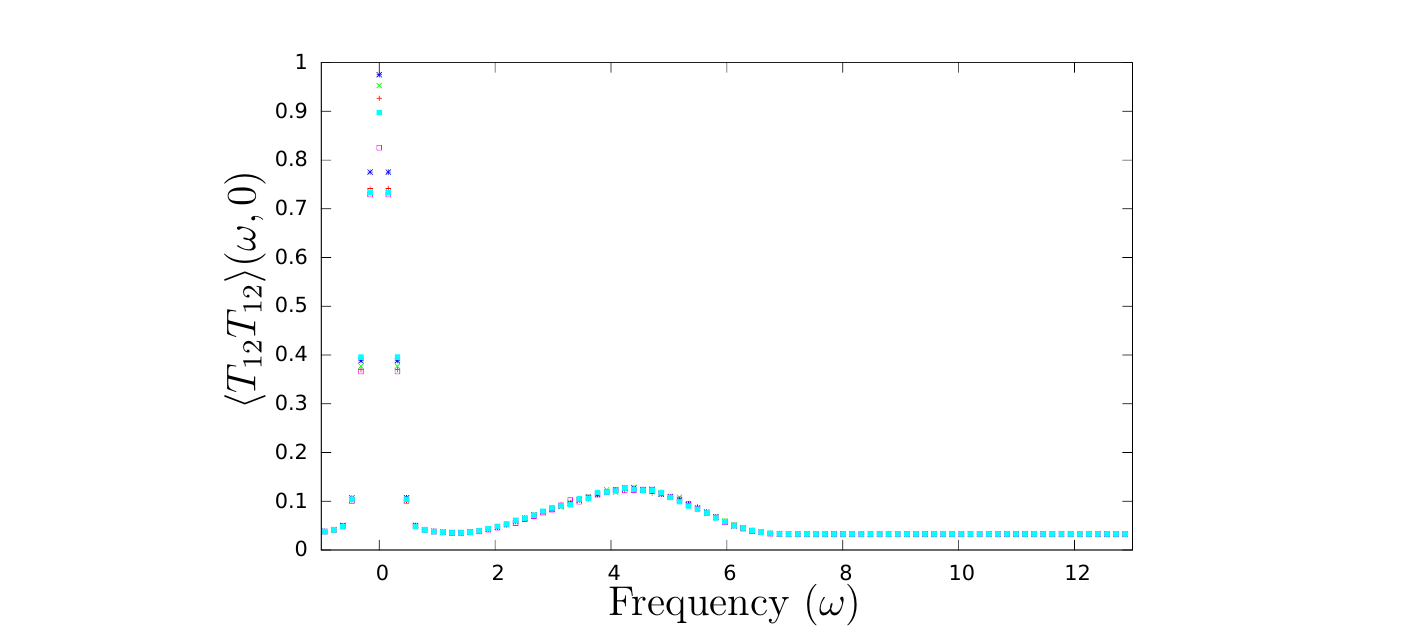}
  \caption{$T_{12}$ correlator in frequency space.}
  \label{fig:viszkow}
\end{figure}

The relevant information, the transport-peak can be extracted from the
Fourier-transformed data shown in Fig.~\ref{fig:viszkow}. To
understand what we see in this figure we recall that in the leading
order of perturbation theory we expect a branch cut starting at $2m$,
and a Dirac-delta peak at $k=0$, just like in the Fourier transform of
$\exvs{\Phi^2(x)\Phi^2(0)}$. The higher order terms result in the
smearing of the cut and the Dirac-delta peak as well, the former
yielding a broad bump, the latter leading to the transport peak. The
desired result is the height of this peak at $\omega=0$.

\begin{figure}[htbp]
  \centering
  \includegraphics[scale=0.7]{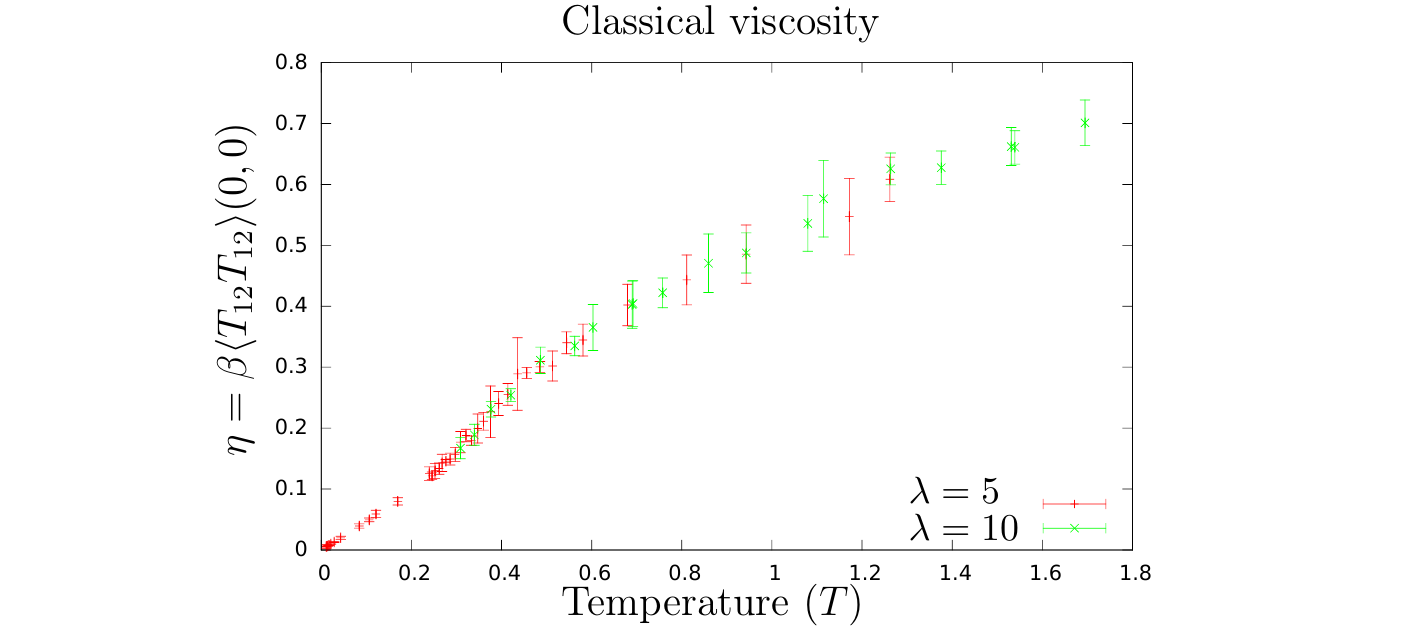}
  \caption{Temperature dependence of the classical shear-viscosity.}
  \label{fig:viszko}
\end{figure}

One can repeat this measurement at different temperatures as well, this
can be seen on Fig.~\ref{fig:viszko}. This figure suggests that the
classical shear viscosity depends more or less linearly on the
temperature. It also follows from dimensional analysis: since
$[\Phi]=1/2$ is the mass dimension of the field, $[T_{12}]\sim 3$, its
correlator has $6$th power of energy dimension. After Fourier
transformation there remains $2$, and after division by the
temperature, there remains $1$, a linear energy dependence. Since the
main source of energy dependence in the classical case is the
temperature, we expect proportionality with $T$.

\begin{figure}[htbp]
  \centering
  \includegraphics[scale=0.7]{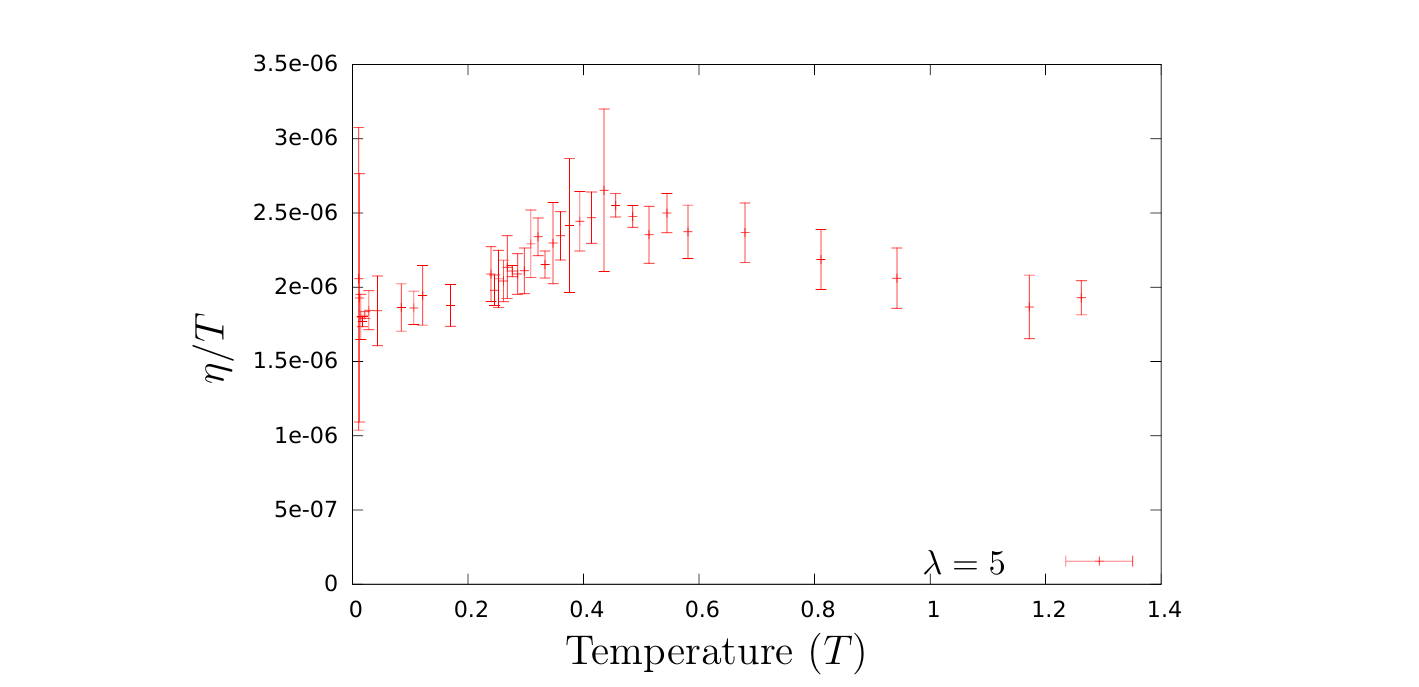}
  \caption{Classical shear viscosity over temperature as a function of
    the temperature. The enhanced behavior is around the phase
    transition point, suggesting that the viscosity also shows a
    critical behavior. }
  \label{fig:viscoalpha}
\end{figure}

To verify numerically this, we also present the $\eta_{cl}/T$ curve in
Fig.\@~\ref{fig:viscoalpha}. We see that the ratio is approximately
constant, but with an enhanced behavior near the critical point. This
figure suggests that the classical viscosity, just like other
susceptibilities, shows a critical behavior, exhibiting a peak at the
phase transition point.

\subsection{Interpretation}

One can compute the classical viscosity in perturbation theory to
leading order, similarly what was done in
Ref.~\cite{Jakovac:1998jd}. With point splitting we can write
\begin{eqnarray}
  S(x)= \exv{T_{12}(x)T_{12}(0)} &&= \lim\limits_{{x'\to x}\atop{y'\to y\to 0}}
  \d_{x1}\d_{x'1}\d_{y1}\d_{y'1} \exv{\Phi(x)\Phi(x')\Phi(y)\Phi(y')}
  =\nn&&=
  \d_1^2 iG_{3D}(x)  \d_2^2 iG_{3D}(x) +   (\d_1\d_2 iG_{3D}(x))^2.
\end{eqnarray}
After Fourier transformation we have
\begin{eqnarray}
  S(p) &&= \int\frac{d^4k}{(2\pi)^4} \left[k_1^2(p-k)_2^2 + k_1 k_2
    (p-k)_1 (p-k)_2 \right]iG_{3D}(k)iG_{3D}(p-k) =\nn&&=
  T^2 \int\frac{d^4k}{(2\pi)^2} \frac{k_1^2(p-k)_2^2 + k_1 k_2
    (p-k)_1 (p-k)_2}{k_0(p_0-k_0)} \rh(k)\rh(p-k).
\end{eqnarray}
This is very similar to the $\Phi^2\Phi^2$ correlation function
discussed in \cite{Jakovac:1998jd}, and also very similar to the
discontinuity of the quantum version of it.

To proceed we use the Green-Kubo relation \eqref{eq:GreenKubo} to
write
\begin{equation}
  \eta_{cl} = 2T \int\!\frac{d^4k}{(2\pi)^4}\, \frac{k_1^2k_2^2}{k_0^2}
  \rh^2(k).
\end{equation}
Evaluating this expression with the free spectral function
\eqref{eq:freespect} we obtain infinity: this means that for free
theories the viscosity, like all other transport coefficient, is
infinite. We can apply a Breit-Wigner approximation for the spectral
function
\begin{equation}
  \rh(k) = \frac{4k_0\gamma_k}{(k_0^2-\omega_k)^2 + 4 k_0^2\gamma_k^2},
\end{equation}
then we can approximate for small width and for $k_0>0$:
\begin{equation}
  \rh^2(k)\approx \frac{2\pi}{\gamma_k\omega_k^2} \delta(k_0-\omega_k).
\end{equation}
This leads to
\begin{equation}
  \eta_{cl} = 4T \int\!\frac{d^3\mathbf{k}}{(2\pi)^3}\,
  \frac{k_1^2k_2^2}{\gamma_k \omega_k^4}.
\end{equation}
This integral is still divergent in the continuum limit. The leading
contribution comes from large momenta. The asymptotics of $\gamma_k
\omega_k$ has been found in \cite{Jakovac:1998jd} (see also
\cite{Aarts:2001yn}), it is $\frac{\lambda^2 T^2}{384\pi}$. In the
remaining $\omega_k^{-3}$ factor one can put $m=0$. Therefore
to leading order we have
\begin{equation}
  \eta_{cl} = \frac{1536 \pi}{\lambda^2 T} \int\!\frac{d^3\mathbf{k}}{(2\pi)^3}\,
  \frac{k_1^2k_2^2}{(k_1^2+k_2^2+k_3^2)^{3/2}} +\dots.
\end{equation}
Rewriting this integral in terms of the dimensionless momenta $ka$,
and integrating it in the Brillouin zone $-\pi< ka<\pi$, one finds
\begin{equation}
  \eta_{cl} = \frac{\eta_{lat}}{a^4},\qquad 
  \eta_{lat} = \frac{321.3\pi}{\lambda^2 T} +\dots.
\end{equation}
The fact that $\eta_{cl}\sim a^{-4}$ is consistent with the results of
\cite{Jakovac:1998jd} for the $\Phi^2$ autocorrelation function. There
a logarithmic divergence was found, therefore the energy-momentum
tensor correlation function, which has two derivatives more than
$\Phi^2$, should scale as $a^{-4}$.

Although this result is just a first order perturbative estimate, the
robust part of the result is that the classical shear viscosity is
proportional to $a^{-4}$. After determining $\eta_{lat}$ by lattice
simulations, this is the way how one can recover the classical value
of the viscosity.

The next question is how can one relate $\eta_{cl}$ to the shear
viscosity of quantum systems. The facit of the comparison of the
classical and quantum calculations \cite{Jakovac:1998jd,Wang:1995qg}
was that the perturbative quantum result was rather close to the
classical one, if one chooses a cutoff $\Lambda\sim T$. The exact
value of the coefficient is not known to be universal, probably it
depends on the quantity in question. But, up to a constant, we can
estimate the result of the full quantum result by setting $\eta \sim
(aT)^4\eta_{cl}$. If we measure the shear viscosity from the lattice,
we have
\begin{equation}
  \eta \sim T^4 \eta_{lat}.
\end{equation}
Below we will use unity for the proportionality constant.

Of course, this formula is based on the assumption that the quantum
result is dominated by the classical fields. Also, unfortunately, we
do not know the coefficient in this formula, nevertheless, we can give
a temperature profile of the $\eta/s$ ratio.

\begin{figure}[htbp]
  \centering
  \includegraphics[scale=0.8]{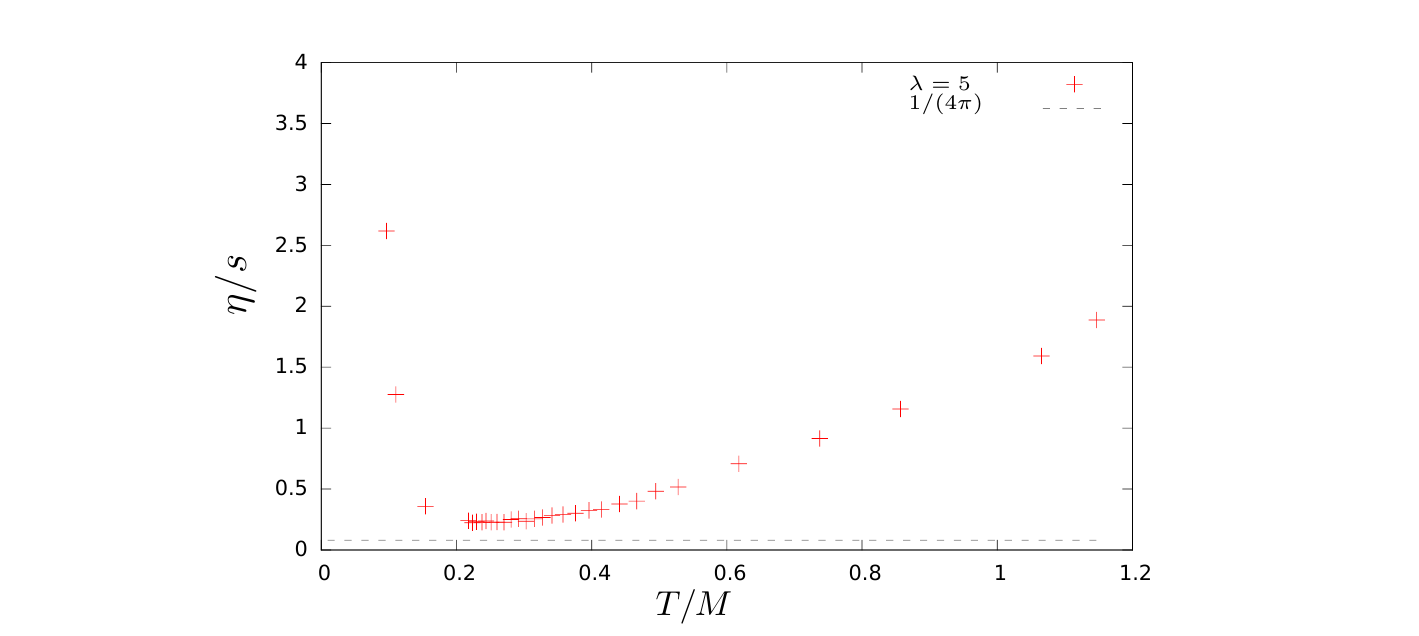}
  \caption{The quantum $\eta/s$ estimated by the classical $\Phi^4$
    theory, using $a=T^{-1}$ lattice spacing.}
  \label{fig:viszko1}
\end{figure}

\begin{figure}[htbp]
  \centering
  \includegraphics[scale=0.8]{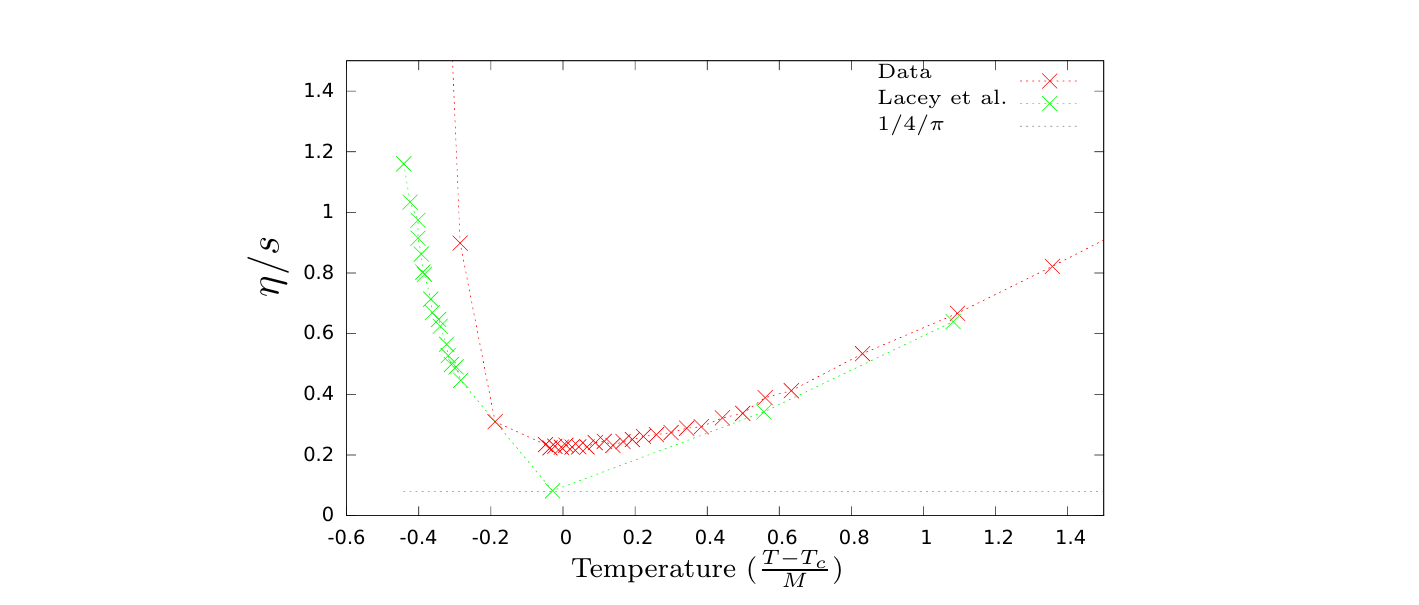}
  \caption{Qualitative comparison of $\eta/s$ of QCD plasma and the
    classical $\Phi^4$ theory, after rescaling. The QCD data are taken
  from \cite{Lacey:2006bc}.}
  \label{fig:compare}
\end{figure}

The entropy density of the quantum system is poorly approximated by
the classical modes, so we use for the $\eta/s$ estimate the entropy
density of a free one component gas with
\begin{equation}
  s = \int\!\frac{d^3\mathbf{k}}{(2\pi)^3}\,\left[\frac{\omega_k}
    {e^{\beta\omega_k}  -1} - \ln(1-e^{-\beta\omega_k}) \right].
\end{equation}
Here $\omega_k^2=\mathbf{k}^2+m_{eff}^2$, where $m_{eff}$ was taken
from the classical simulations. In this way we can present our
estimate for the $\eta/s$ ratio in Fig.~\ref{fig:viszko1}.
We see a curve typical for the behavior of the shear viscosity in any
matter near the phase transition point. For a qualitative comparison
we show the $\eta/s$ ratio for QCD \cite{Lacey:2006bc} in
Fig.~\ref{fig:compare} versus our results that was rescaled to fit to
the high temperature part.

\section{Summary}

In this work we used numerical simulations of the quartic classical
field theory to give an estimate for the shear viscosity. To this end
we solved the discretized classical equations of motion. This leads to
thermalization, where one can determine the expectation value of
different observables by time averaging. We first determined the field
autocorrelation function, and analyzed it by assuming quasiparticle
behavior. Then we measured the correlator of the $12$ component of the
energy-momentum tensor. Using the Green-Kubo formula this quantity is
proportional to the shear viscosity. The shear viscosity $\eta_{lat}$
was found to depend approximately linearly on $T$ which was expected
by classical dimensional analysis. This behaviour was superimposed by
a characteristic critical behavior near the phase transition
regime. Finally we pointed out that the classical viscosity comes from
the lattice viscosity as $\eta_{cl}= \eta_{lat} a^{-4}$ where $a$ is
the lattice spacing. Translating the classical result into the shear
viscosity of the quantum system, we argued that $a^{-1}\sim T$ is the
correct choice, but the proportionality constant is not known. This
allows us to make an estimate also on the temperature profile of
$\eta/s$, which turns out to be rather similar to the result of QCD
near the critical region.

As future prospects we plan to repeat this analysis to other models
including gauge theories. Another interesting extension could be to
study the effects of the quantum corrections to the equations of
motion. This could give a hint on the reliability of the classical
estimate.

\section*{Acknowledgement}

The authors acknowledge useful discussions with A. Patk\'os,
Zs. Sz\'ep. This research was supported by the grant K-104292 from the
Hungarian Research Fund (OTKA).

\end{document}